\documentstyle[twocolumn,prl,aps,epsf]{revtex}
\begin{document}

\title{Possible Triplet Electron Pairing and an Anisotropic Spin Susceptibility in Organic Superconductors (TMTSF)$_2$X}

\author{A. G. Lebed$^{1,2}$, K. Machida$^{1}$ and M. Ozaki$^{3}$}

\address{$^1$Department of Physics, Faculty of Science, Okayama University, Okayama, Japan}
\address{$^2$Landau Institute for Theoretical Physics, 2 Kosygina
Street, Moscow,  Russia}
\address{$^3$Department of Physics, Kochi University, Kochi,  Japan}

\maketitle

\abstract
{We argue that (TMTSF)$_2$PF$_6$ compound under
 pressure is likely a triplet superconductor with a vector order parameter ${\bf d}({\bf k}) \equiv (d_a({\bf k}) \neq 0, \ d_c({\bf k}) = ?, \ d_{b'}({\bf k}) = 0)$; $|d_a({\bf k})| > |d_c({\bf k})|$. 
It corresponds to an anisotropic spin susceptibility at $T=0$: $\chi_{b'} = \chi_0$, $\chi_a \ll \chi_0$, where $\chi_0$ is its value in a metallic phase.
[The spin quantization axis, ${\bf z}$, is parallel to a so-called
 ${\bf b'}$-axis]. 
We show that the suggested order parameter explains why the upper critical field along the ${\bf b'}$-axis exceeds all paramagnetic limiting fields, including that for a nonuniform superconducting state, whereas the upper critical field along the ${\bf a}$-axis (${\bf a} \perp {\bf b'}$) is limited by the Pauli paramagnetic effects [I. J. Lee, M. J. Naughton, G. M. Danner and P. M. Chaikin, Phys. Rev. Lett. ${\bf 78}$, 3555 (1997)].
The triplet order parameter is in agreement with the recent Knight shift measurements by I. J. Lee et al. as well as with the early results on a destruction of superconductivity  by nonmagnetic impurities and on the absence of the Hebel-Slichter peak in the NMR relaxation rate. 

 }

\vspace{1cm}

Quasi-one-dimensional (Q1D)
organic compounds \\
(TMTSF)$_2$X  (X = PF$_6$, ClO$_4$, etc.) 
have been intensively
investigated since the discovery of  
superconductivity$^{1,2}$ in the first organic 
superconductor (TMTSF)$_2$PF$_6$. 
From the beginning, it was clear that their properties are unusual. 
It was found$^{3-8}$ that superconductivity in (TMTSF)$_2$X (X = PF$_6$, ClO$_4$) is destroyed by nonmagnetic impurities. 
This was interpreted in terms of a possible triplet pairing
 of electrons$^9$. 
Another unusual feature, the absence of the Hebel-Slichter peak
 in the $1/T_1$ NMR data in (TMTSF)$_2$X (X = PF$_6$, ClO$_4$)$^{10,11,12}$, was prescribed$^{13}$ to the existence of zeros of a superconducting order parameter on the Q1D Fermi surfaces (FS). 
As was stressed$^{13}$, the early experiments$^{3-8, 10,11}$ provided information only about an orbital part of the order parameter and could not distinguish between some triplet and singlet pairings$^{2,13}$. 

To reveal triplet superconductivity, experimental tests which probe a spin part of an order parameter are essential.
Among them, are: a surviving of triplet superconductivity in Q1D
 case$^{14-17}$ at magnetic fields higher than both the upper orbital critical field and the Clogston paramagnetic limit$^{18}$, observation of spin-wave exitations$^{15}$, the Knight shift measurements$^{12}$ and some others.
Nowadays, interest in a possible triplet pairing has been renewed due to remarkable measurements of the upper critical  fields (which are sensitive to a spin part of the order parameter) in (TMTSF)$_2$ClO$_4$ and in (TMTSF)$_2$PF$_6$ at
 $P \simeq 6 \ kbar$ by Naughton, Lee, Chaikin and Danner$^{19-21}$ and due to the theoretical analysis$^{16}$ of these experiments. 
The experimental fields along ${\bf b'}$-axis (which are 3 times
 bigger$^{20,21}$ than the Clogston paramagnetic limit) were shown$^{16}$ to be even bigger than the paramagnetic limit$^{16,22}$ for the Larkin-Ovchinnikov-Fulde-Ferrell (LOFF) phase$^{23}$. Therefore, measurements$^{19-21}$ were interpreted$^{16,19-21}$ in term of triplet superconductivity.
Recently, Lee et al.$^{12}$ have found no change of the Knight shift for ${\bf H} \parallel {\bf b'}$ in a superconducting phase of (TMTSF)$_2$PF$_6$ at $P \simeq 6 \ kbar$. This is consistent with the results$^{16,19-21}$ and strongly supports the triplet scenario$^{9, 16,19-21}$ of superconductivity.

The goals of our paper are:
1) To calculate the paramagnetic limited field along ${\bf b'}$-axis,
 $H^{b'}_p$, for the LOFF phase in a Q1D superconductor, taking account of both the paramagnetic$^{16}$ and orbital destructive effects against superconductivity.
(We show that the calculated value of $H^{b'}_p$ is 4-5 times less
 than the experimental fields$^{20,21}$ in (TMTSF)$_2$PF$_6$); 
2) To demonstrate that the value of $H^{b'}$ becomes consistent
with$^{20,21}$ if we switch off the paramagnetic effects.
(These indicate that an electron spin susceptibility along
 ${\bf b'}$-axis, $\chi_{b'}$, at $T=0$ is equal to its value in a metallic state, $\chi_0$, which is a distinct feature of triplet superconductivity$^{24,27}$);
3) To stress that the experimental critical fields$^{20,21}$ along
 the conducting chains (i.e., along ${\bf a}$-axis), $H^a_p$, are strongly paramagnetically limited and thus the corresponding electron spin susceptibility $\chi_a \ll \chi_0$ at $T=0$;  
4) To show that the above described properties are naturally
 explained within the framework of a triplet superconductivity scenario with the following vector order parameter frozen into the crystalline lattice  (i.e., the case of strong spin-orbit coupling$^{27}$):
\begin{equation} 
{\bf d}({\bf k}) = (d_a({\bf k})  \neq 0, d_c({\bf k}) = ?, 
d_{b'}({\bf k}) = 0);  |d_a({\bf k})| > |d_c({\bf k})|
\end{equation}
 corresponding to the BCS-pair's wave function
\begin{equation}
\Psi ({\bf k}) =  [-d_a({\bf k}) + i d_c({\bf k})] \ | \uparrow \ \uparrow \bigl>  +  [d_a({\bf k}) + i d_c({\bf k})]  \ | \downarrow \ \downarrow \bigl>
\end{equation}
and to the anisotropic spin susceptibility at $T = 0$:
\begin{equation}
\chi_{b'} = \chi_0 \ \ , \ \ \chi_a \ll \chi_0 \ \ ,
\end{equation}
where $| \uparrow \bigl>$ $(| \downarrow \bigl>)$ stands for 
a spin-up (spin-down) electron with respect to the quantization axis
 ${\bf z} \parallel {\bf b'}$ [${\bf a(x)} \perp
 {\bf b'(z)} \perp {\bf c^*(y)}$], the momentum ${\bf k}$ defines the position on the FS. [We stress that ${\bf b'}$ is the easy axis for a spin direction in a spin-density-wave (SDW) phase of (TMTSF)$_2$PF$_6$. Thus, one may expect that the order parameter (1) is the most stable since it corresponds to the BCS pairs (2) only with $S_{b'} \equiv S_z= \pm 1$]. At the end of the paper, we discuss some consequences of a group theory classification of the possible triplet phases, including the most probable orbital part of the order parameter and a possibility to break the time reversal symmetry. 

Q1D electron spectrum corresponds to two open sheets of the FS$^{1,2}$:
\begin{equation}
\epsilon^\pm ({\bf p}) = \pm v_F \ (p_a \mp p_F) - 2 t_b
 \cos(p_b b') - 2t_{c}\cos(p_c c^*)  \ \  ,
\end{equation}
where $+(-)$ stands for the right (left) sheet of the FS; $v_F = t_a a/\sqrt{2}$ and
 $p_F$  are the Fermi velocity and Fermi momentum, respectively; $t_a \simeq 1600 \ K$, $t_b \simeq 200 \ K$ and $t_c \simeq 5 \ K$; ($\hbar =1$).

Singlet (S = 0) and triplet (S = 1) phases are characterized by the
following wave functions of the BCS pair's$^{27}$:
\begin{equation}
\psi_s({\bf k}, {\bf r}) = ( \ | \uparrow \ \downarrow \bigl> \ - \ |
\downarrow \ \uparrow \bigl> \ ) \ \psi({\bf k}, {\bf r}) \ , \ \ \ \
 S=0 \ ;
\end{equation} 
\begin{eqnarray}
\psi_t({\bf k}, {\bf r}) = && | \uparrow \uparrow \bigl>  
[- d_x ({\bf k , {\bf r}}) + id_y ({\bf k}, {\bf r})] + 
( |\uparrow\downarrow \bigl>+|\downarrow\uparrow \bigl>) 
\nonumber \\
&&d_z ({\bf k}, {\bf r}) +|\downarrow\downarrow \bigl> [d_x ({\bf k}, {\bf r}) + i d_y ({\bf k}, {\bf r})], \ \ S=1.
\end{eqnarray}
[In Eqs. (5,6), $S$ is the total spin of the BCS pair, ${\bf r}$ is its coordinate of a center of masses; $\psi_s({\bf k}, {\bf r}) =\psi_s({-\bf k}, {\bf r})$, ${\bf d({\bf k}, {\bf r}}) 
= - {\bf d({\bf -k}, {\bf r}})]$.  $\\$
 At $H \rightarrow 0$, $\psi({\bf k}, {\bf r})$ and ${\bf d({\bf k}, 
{\bf r})}$ do not depend on ${\bf r}$. Electron spin susceptibility tensor, $\chi_{i,j}$,  at $T = 0$ for a singlet phase is $\chi_{i,j} =  0$ whereas for a triplet phase is given by$^{27}$ :
\begin{equation}
\chi_{i,j} =  \chi_0 \ \biggl< \delta_{i,j} - \frac{ d^*_i({\bf k}) d_j({\bf k}) }{ {\bf d^*({\bf k})} {\bf d({\bf k})} } \biggl>_{\bf k} \ , 
\end{equation}
where $\delta_{i,j} =1$ if $i=j$ and $\delta_{i,j} =0$ if $i \neq j$; 
 $\bigl< |{\bf d({\bf k})}|^2 \bigl>_{\bf k}=1$,
 $\bigl<...\bigl>_{\bf k}$ means an averaging over the FS. [Here, we consider only unitary triplet phases$^{27}$ (i.e., $d_a({\bf k})d^*_c({\bf k}) = d^*_a({\bf k})d_c({\bf k})$)].

\begin{figure}
\begin{center}
\leavevmode
\epsfxsize=80mm
\epsfbox{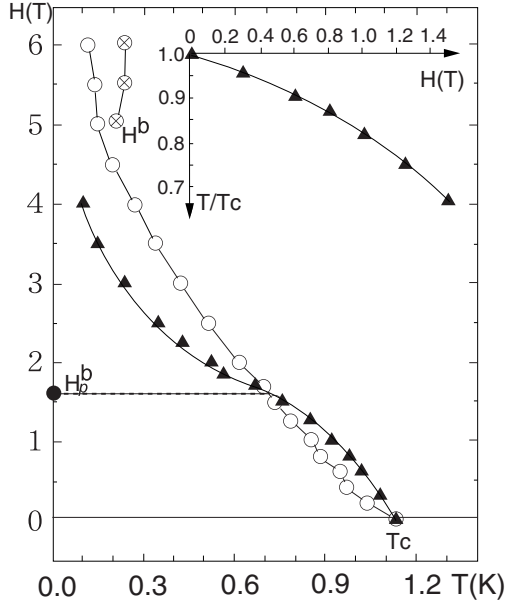}
\vspace{5mm}
\end{center}
\caption{
Circles stand for the critical magnetic fields along
${\bf b'}$-axis:
open circles show experimental curve$^{20,21}$, 
a full circle corresponds to the calculated paramagnetically 
limited value of $H^b_p$ at $T=0$ in a singlet superconductor 
whereas crossed circles show the calculated non-paramagnetically 
limited critical fields $H^b(T)$ for a triplet order parameter (1). 
Triangles stand for the experimental critical fields$^{20,21}$ 
along ${\bf a}$-axis, $H^a_p(T)$, (${\bf a} \perp {\bf b'}$). 
In the inset, the experimental values$^{20,21}$ of $H^a_p(T)$ 
are shown in comparison with the calculated paramagnetically 
limited field (full line) for a triplet order parameter (1) (see the text). 
$\\$
}
\label{fig:Fig1}
\end{figure}

At first we consider the case ${\bf H} \parallel {\bf b'} ({\bf z})$. 
In singlet phase (5), superconductivity is destroyed by paramagnetic effects in arbitrary directed magnetic field. 
In a triplet phase (6), as it follows from Eq. (7), $d_{b'}({\bf k}) \equiv d_z({\bf k})$ component is responsible for the deviation of the spin susceptibility $\chi_{b'} \equiv \chi_{zz}$ from $\chi_0$. 
If $d_{b'}({\bf k}) \neq 0$ there exist two related phenomena: the paramagnetic destructive mechanism against superconductivity and a change of the Knight shift at $T < T_c(H)$. 
Let us calculate the upper critical field for ${\bf H} \parallel {\bf b'}$.  By using a common approach$^{28}$ to the upper critical field of a clean superconductor$^{25}$ with open electron orbits and with one-component order parameter, it is possible to prove that Eq. (5) of Ref. [28],
\begin{eqnarray}
\Delta(x) = &&\frac{g}{2}\int_{ |x-x_1| > d} 
\frac{2 \pi T dx_1 }{v_F \sinh(\frac{2 \pi T |x-x_1|}{ v_F})}J_0\biggl[\frac{2 \alpha  \mu_B H (x-x_1) S_z}{v_F}\biggl]
\nonumber\\
&&\times J_0\biggl(2\lambda \sin \biggl[\frac{ \omega_c(x-x_1)}{2v_F}\biggl]\sin \biggl[\frac{ \omega_c(x+x_1)}{2v_F}\biggl]\biggl)
\nonumber\\
&&\times\cos
\biggl[\frac{2 \mu_B H (x-x_1) S_z}{v_F} \biggl] \ \Delta(x_1),
\end{eqnarray}
is extended to a singlet phase $ \psi_s({\bf k},{\bf r}) \equiv f({\bf k}) \Delta(x) $ as well as to the triplet phases $ {\bf d_1}({\bf k}, {\bf r}) \equiv  (d_a =1, d_c =0, d_{b'} =0) f({\bf k})\Delta(x)$ and 
${\bf d_2}({\bf k}, {\bf r}) \equiv  (d_a =0, d_c =0, d_{b'} =1) f({\bf k})\Delta(x)$. 
[Here, $ \bigl< |f({\bf k})|^2 \bigl>_{\bf k}=1$; $g$ is an effective electron interaction constant, $d$ is a cutoff distance; $\alpha = \sqrt{2} t_b/t_a$, $\omega_c = e v_F H c^*/c$, $\lambda = 4 t_{c}/\omega_c$; $\mu_B$ is a Borh magneton, $ e$ and $c$ are the electron charge and the velocity of light, correspondingly; $ S_z= 1$ for singlet and for ${\bf d}_{2}$-triplet phases whereas $S_z = 0$ for ${\bf d}_1$-triplet phase.  
By solving Eq. (8) numerically for $S_z =1$, $\alpha = 0.17$, 
$|dH^{b'}/dT|_{Tc} \simeq 2 \ T/K$,
$v_F = 10^{7} cm/sec, \ t_c \simeq 3 K, \ T_c(0) =1.14 \ K, \ c^* = 13.6 \ A$ (see Refs. [1,2,19-21,29]), we found that the calculated value of the paramagnetic limited critical field, $H_p^{b'} \simeq 1.3-1.4 \ T$, is 4-5 times less than the experimental ones$^{20-21}$ (see Fig. \ref{fig:Fig1}). 
A similar analysis for ${\bf d_1}$-triplet phase (which is not paramagnetically limited) shows
 that superconductivity survives at $H^{b'} \simeq 6 \ T$ and $T \simeq 0.2-0.25 \ K$ in qualitative agreement with experiments$^{20,21}$ (see Fig. \ref{fig:Fig1}). 
On the basis of the calculation of $H_p^{b'}$ and $H^{b'}$, we can conclude that $|d_{b'}({\bf k})| \equiv |d_z({\bf k})|\simeq 0$ in Eq. (7) and thus $\chi_{b'} \equiv \chi_{zz} \simeq \chi_0$.  Note that the recent Knight shift measurements$^{12}$ are also in favor of $\chi_{b'} = \chi_0$ below $T_c(H)$. 

If we consider the case ${\bf H} \parallel {\bf a} ({\bf x})$ then $d_a \equiv d_x$-component of the order parameter (6) is responsible for the destructive paramagnetic effects against superconductivity and for the change of the Knight shift at $T < T_c(H)$ (see Eq. (7)).
Let us calculate the critical field for ${\bf H} \parallel {\bf a}$ in ${\bf d_1}({\bf r}) \equiv (d_a \neq 0, d_c \neq 0, d_{b'} =0) \ \Delta(x)$
triplet phase (which is paramagnetically limited). The corresponding linearized gap equation can be
 obtained from the common Eq. (5) of Ref. [28]:
\begin{eqnarray}
\Delta(x) =&& \frac{g}{2} \int^{2 \pi}_0 \frac{d \phi}{2 \pi}
\int^{\infty}_{ |x-x_1| > \sqrt{2} d  |\sin \phi| / \gamma}
\nonumber\\
&&\Delta( x_1) \ \frac{\sqrt{2} \gamma \pi T dx_1}{v_F \sin \phi \sinh[\frac{\sqrt{2} \gamma \pi T |x-x_1|}
{v_F \sin \phi}]}
\nonumber\\
&&\times J_0 \biggl( \frac{\sqrt{2} \lambda \gamma}{\sin \phi}
  \sin[\frac{ \omega_c (x-x_1)}{2v_F}]
\sin[\frac{ \omega_c (x+x_1)}{2v_F}] \biggl) 
\nonumber\\
&&\times\cos\biggl[\frac{\sqrt{2} \gamma \mu_B H S_z (x-x_1)}{v_F \sin \phi}\biggl],
\end{eqnarray}
where $\gamma = t_a a/(2t_b b)$.
Numerical solution of Eq. (9) (with the same values of parameters as Eq. (8)) shows that the best fitting of the data$^{20,21}$ at $H \leq 1.5 \ T$ (see Fig. \ref{fig:Fig1}) corresponds to $S_z  \simeq 0.9$ (i.e., $d_a \simeq 0.9$, $\chi_a \simeq 0.2 \ \chi_0 \ll \chi_0$) and $|dH^a/dT|_{Tc} \simeq 8 \ T/K$. 
The latter is in a good agreement with the experimental slopes$^{20,21}$ $|dH^{b'}/dT|_{Tc} \simeq 2 \ T/K$ since the value of $t_b/t_a \simeq 8.5$ is known$^{29}$. Note that the accuracy of our calculations does not allow us to distinguish between the triplet phases with $d_c = 0$ and $|d_a| > |d_c|$.

Summarizing, our analysis of the experimental critical fields$^{20,21}$ measured in (TMTSF)$_2$PF$_6$ at $P \simeq 6 \ kbar$ has shown that paramagnetic destructive effects against superconductivity do not affect $H^{b'}$ whereas $H^a$ is paramagnetically limited at $H  \leq 1.5 \ T$. 
These are naturally explained within a triplet scenario of
 superconductivity$^{9, 16, 19-21}$ with the triplet order parameter (1).  
We suggest to measure the Knight shift along the ${\bf a}$-axis at $H \leq 1.5 \ T$ and $T < T_c(H)$ to prove the order parameter (1). 
Note that temperature dependence of the critical field along a-axis, $H^a(T)$, changes drastically$^{20,21}$ at $H \geq1.5 \ T$.  
We speculate that at $H \geq1.5 \ T$ there may appear a triplet phase with ${\bf d}({\bf k}) \perp {\bf H}$, which minimizes the magnetic contribution to the free energy$^{30}$. Nevertheless, we cannot completely exclude another possibility - the appearance of the LOFF state at $H \geq1.5 \ T$ for ${\bf H} \parallel {\bf a}$.  
Note that our theoretical analysis of the critical fields is based on the the Fermi-liquid picture$^{29}$ proved at $P \simeq 6 \ kbar$ in (TMTSF)$_2$PF$_6$. At higher pressures, $P \simeq 9.8 \ kbar$, the behavior of (TMTSF)$_2$PF$_6$ may deviate from the Fermi liquid one$^{31}$. 

At the end of the paper, we would like to make a few comments based on symmetry arguments.
We classify the possible triplet phases in the case of strong spin-orbit coupling for orthorhombic (D$_{2h}$) and triclinic (C$_{i}$) point group symmetries 
(see Table \ref{table:Table1}), where
the matrix order parameter ${\hat \Delta}({\bf k}) = 
d_i({\bf k}) {\hat \tau}_i$, (${\hat \tau}_i = i {\hat \sigma}_i {\hat \sigma_y}$; ${\hat \sigma}_i$ are the Pauli matrices). 
As it seen from Table \ref{table:Table1}, there are no degenerated orbital states, 
thus a time reversal symmetry is broken 
only if a nonunitary triplet phase appears$^{27}$. 
In our particular case, this happens when $d_a({\bf k})d^*_c({\bf k}) \neq d^*_a({\bf k})d_c({\bf k})$.
Using the expression  for a gap in a quasi-particle spectrum$^{27}$, $\delta(k)$ = $|{\bf d(k)}|$ (the unitary case), it is possible to make sure that there are no generic phases with the lines of zeros on the FS in accordance with a common theorem$^{32}$. 
This is in agreement with the experimental data$^{26, 33}$ which seem to be in favor of fully gapped FS and against the existence of isolated zeros on the FS$^{32}$. 
Therefore, we speculate that the orbital part of the order parameter is likely $d_a({\bf k}) \sim d_c({\bf k}) \sim sgn(k_a)$ which corresponds to a fully gapped Q1D sheets of the FS. From Table \ref{table:Table1}, it is possible to conclude that, for a triclinic space group of (TMTSF)$_2$PF$_6$, the most generic case is
$d_a \neq 0$, $d_c \neq 0$ and $d_{b'} \neq 0$. 
However, it is known$^{1,2,34}$ that the spin dependent interactions in a SDW phase of (TMTSF)$_2$PF$_6$ (which has a common boundary with the superconducting phase) result in an alignment of spins along ${\bf b'}$-axis. Therefore, it is natural to expect the form (1) for the superconducting order parameter corresponding to the absence of the BCS pairs with $S_{b'} = 0$ (see Eq. (2)). 

One of us (A.G.L.) is thankful to Agterberg, N. N. Bagmet, K. Behnia, S. Brown, E. V. Brusse, P. M. Chaikin,  T. Ishiguro, H. Fukuyama, I.J. Lee, P. Lee, Y. Maeno, V. P. Mineev, M. J. Naughton, K. Oshima, M. Sigrist, V. M. Yakovenko for useful discussions. A.G.L. is especially thankful to S. Brown, P. M. Chaikin, I. J. Lee and M. J. Naughton
for fruitful and numerous discussions during a workshop organized by M. J. Naughton.

\begin{table}
\caption{Triplet order parameter ${ \hat \Delta}(k)$ 
for $D_{2h}$ and $C_{i}$ groups. (A$\sim$I are constants)}
\begin{tabular}{ccc}
\hline
Group&Representation&Order parameter ${\hat \Delta}(k)$\\
\hline
$D_{2h}$&$A_{1u}$&A$k_x{\hat \tau_x}$+B$k_y{\hat \tau_y}$+C$k_z{\hat \tau_z}$\\
&$B_{1u}$&A$k_x{\hat \tau_y}$+B$k_y{\hat \tau_x}$\\
&$B_{2u}$&A$k_x{\hat \tau_z}$+B$k_z{\hat \tau_x}$\\
&$B_{3u}$&A$k_y{\hat \tau_z}$+B$k_z{\hat \tau_y}$\\
\hline
$C_{i}$&$A_{u}$&A$k_x{\hat \tau_x}$+B$k_y{\hat \tau_y}$+C$k_z{\hat \tau_z}$+D$k_y{\hat \tau_x}$\\
&&+F$k_x{\hat \tau_z}$+G$k_z{\hat \tau_x}$+H$k_y{\hat \tau_z}$+I$k_z{\hat \tau_y}$\\
\hline
\end{tabular}
\label{table:Table1}
\end{table}


%
\pagebreak


\begin{references}
\bibitem{LeeHo} The other effects which can increase the paramagnetic limits seem to be weak. Indeed, spin-orbital scattering is not significant$^{20,21,25}$; the specific heat jump is close to its week coupling BCS value$^{26}$.
\end{references}
\end{document}